\documentclass[aps, prb, superscriptaddress, twocolumn, showpacs, floatfix ]{revtex4-1}

\usepackage{graphicx}  
\usepackage{dcolumn}   
\usepackage{bm}        
\usepackage{amssymb}   

\usepackage{pdfpages}
\usepackage{amsfonts}
\usepackage{amsmath}
\usepackage{color}
\usepackage{textcomp} 
\usepackage{dcolumn}
\usepackage{microtype}   
\usepackage{mathtools}

\newcommand{\newc}{\newcommand}

\newc{\beq}{\begin{equation}}
\newc{\eeq}{\end{equation}}
\newc{\kt}{\rangle}
\newc{\br}{\langle}
\newc{\beqa}{\begin{eqnarray}}
\newc{\eeqa}{\end{eqnarray}}
\newc{\pr}{\prime}
\newc{\longra}{\longrightarrow}
\newc{\ot}{\otimes}
\newc{\rarrow}{\rightarrow}
\newc{\h}{\hat}
\newc{\bom}{\boldmath}
\newc{\btd}{\bigtriangledown}
\newc{\al}{\alpha}
\newc{\be}{\beta}
\newc{\ld}{\lambda}
\newc{\sg}{\sigma}
\newc{\p}{\psi}
\newc{\eps}{\epsilon}
\newc{\om}{\omega}
\newc{\mb}{\mbox}
\newc{\tm}{\times}
\newc{\hu}{\hat{u}}
\newc{\hv}{\hat{v}}
\newc{\hk}{\hat{K}}
\newc{\ra}{\rightarrow}
\newc{\non}{\nonumber}
\newc{\hs}{\hspace}
\newc{\longla}{\longleftarrow}
\newc{\ts}{\textstyle}
\newc{\f}{\frac}
\newc{\df}{\dfrac}
\newc{\ovl}{\overline}
\newc{\bc}{\begin{center}}
\newc{\ec}{\end{center}}
\newc{\dg}{\dagger}
\newc{\prh}{\mbox{PR}_H}
\newc{\prq}{\mbox{PR}_q}
\newc{\tr}{\mbox{Tr}}
\newc{\pd}{\partial}
\newc{\qv}{\vec{q}}
\newc{\pv}{\vec{p}}
\newc{\dqv}{\delta\vec{q}}
\newc{\dpv}{\delta\vec{p}}
\newc{\mbq}{\mathbf{q}}
\newc{\mbqp}{\mathbf{q'}}
\newc{\mbpp}{\mathbf{p'}}
\newc{\mbp}{\mathbf{p}}
\newc{\mbn}{\mathbf{\nabla}}
\newc{\dmbq}{\delta \mbq}
\newc{\dmbp}{\delta \mbp}
\newc{\T}{\mathsf{T}}
\newc{\J}{\mathsf{J}}
\newc{\sfL}{\mathsf{L}}
\newc{\C}{\mathsf{C}}
\newc{\B}{\mathsf{M}}
\newc{\V}{\mathsf{V}}

\newc{\Cn}{\mathcal{C}}
\newcommand{\eq}[1]{\begin{align}#1\end{align}}

\begin{document}

\title{Local entanglement structure across a many-body localization transition}
\author{Soumya Bera}
\affiliation{Max-Planck-Institut f\"ur Physik komplexer Systeme, 01187 Dresden, Germany}
\author{Arul Lakshminarayan}
\affiliation{Department of Physics, Indian Institute of Technology Madras, Chennai 600036, India}
\affiliation{Max-Planck-Institut f\"ur Physik komplexer Systeme, 01187 Dresden, Germany}
\date{\today}

\begin{abstract}
Local entanglement between pairs of spins, as measured by concurrence, is investigated in a disordered spin 
model that displays a transition from an ergodic to a many-body localized phase in excited states. It is 
shown that the concurrence vanishes in the ergodic phase and becomes nonzero and increases in the many-body 
localized phase. This happen to be correlated with the transition in the spectral statistics from 
Wigner to Poissonian distribution. 
A scaling form is found to exist in the second derivative of the concurrence with the 
disorder strength. It also displays a critical value for the localization transition 
that is close to what is known in the literature from other measures. An exponential decay of concurrence 
with distance between spins is observed in the localized phase. Nearest neighbor spin concurrence in this 
phase is also found to be strongly correlated with the disorder configuration of onsite fields: nearly similar 
fields implying larger entanglement.
\end{abstract}


\maketitle
\section{Introduction}
Anderson localization in non-interacting one-particle states has now long-been studied  and is of special 
interest in the context of low-temperature physics and metal-insulator transitions~\cite{kramer93,evers08}. 
Localization in low dimensional interacting disordered quantum systems, in particular spin chains and 
spinless fermions, are  currently under active investigation~\cite{BAA, Mirlin,oganesyan07, nandkishore15, 
altman_rev15, Znidaric:2008cr,pal10, Berkelbach:2010ib,Monthus:2010gd,bardarson12, IyerPRB15, 
Huse:2013bw, Serbyn:2013he, Bauer:2013jw, Imbrie:2014vo, Vosk:2014jl,Pekker:2014bj,kjaell14, 
Laumann:2014ju,Andraschko:2014bw,chandran14,DeLuca:2014ch, luitz15, Lazarides2014, Ponte2015, Agarwal:2015cu, 
bera15, sonika15,Davakul15,PixleyPRL15, ModakPRL15}. The issue of ``eigenstate thermalization", and a high 
temperature quantum phase transition to the many body localized (MBL) phase is of special interest. Many spin 
models have been studied in this context as well as spinless interacting fermion models, the two being 
related, at least in one-dimension, by the Jordan-Wigner transform. Recently experimental evidence of a MBL 
phase has also been observed in cold atom systems~\cite{schreiber15, schreiber15_v2}, amorphous 
indium-oxide~\cite{Ovadia2015} as well as in a ion-trap experiment~\cite{MonroeMBL2015}.

Interacting quantum systems can be characterized by non-integrability and may have properties of 
random states, that are such that typical subsystems behave as if the rest of the system is a thermal 
bath~\cite{nandkishore15}. That a single many-body excited eigenstate may have this property is referred to as the 
Eigenstate Thermalization Hypothesis (ETH)~\cite{Deutsch:1991, Srednicki:1994, Srednicki99, Rigol:2008bf}. 
Random states in this context are those whose components in a generic basis are uniformly distributed given 
only the constraint of normalization, such as the eigenstates of random matrices from the classical Gaussian 
ensembles~\cite{Mehtabook}. Low dimensional quantum chaotic systems, such as coupled quartic oscillators, 
two-dimensional quantum billiards, Hydrogen atom in a strong magnetic field  etc., are 
well-known examples of this~\cite{LesHouches91}. In the many-body context even one-dimensional spin models 
without disorder can have such eigenstates, for example the Ising model with both a transverse and a 
longitudinal magnetic field is one such~\cite{Karthik07,KimHuse13}. Another well-studied system in which the ETH is 
obtained is the Bose-Hubbard model~\cite{KollathPRL07, Trotzky2012, SorgPRA14, DimaBH15}.

\begin{figure}[tphb]
\centering
\includegraphics[width=0.975\columnwidth]{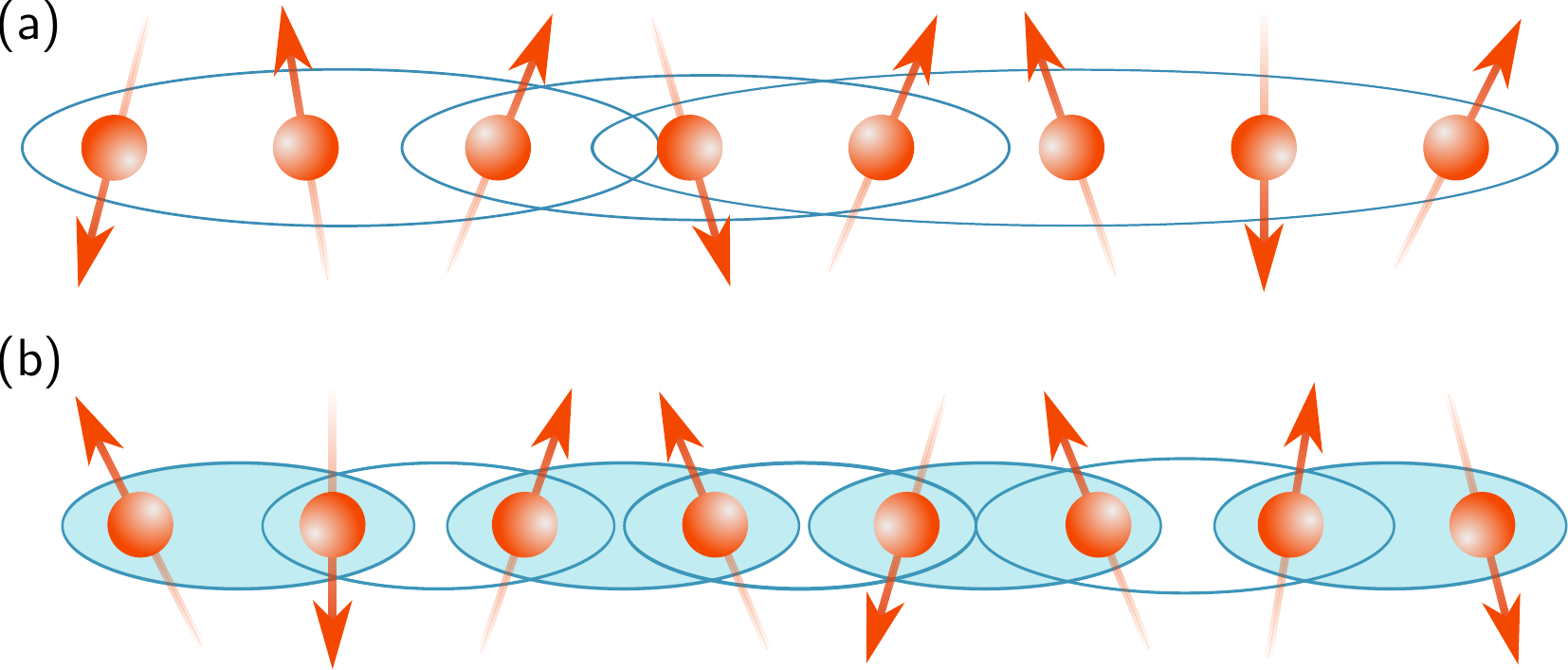}
\caption{Illustration of local entanglement structure in ETH~(a) and many-body localized phase~(b) of a random spin 
chain. In the ETH phase 
the entanglement is shared among all spins, which give rise to volume law bipartite entanglement. 
In contrast many-body localized phase has an area law entanglement and a large part of entanglement 
is stored in nearest neighbor spins.}
\label{fig:SpinEntg}
\end{figure}
These excited states are characterized by large entanglement between subsystems. In fact, the entanglement 
is typically nearly the maximum possible bipartite entanglement, that goes as $\log N$ where $N$ is the 
dimensionality of the Hilbert space of the smaller subsystem.  

This is then proportional to the number of particles in the 
subsystem and this extensive situation is referred as the volume law~\cite{Grover:2014wm, kjaell14, 
luitz15}. In this regime the eigenstates have large multipartite entanglement and we can expect that the 
entanglement among small subsystems is absent~(see Fig.~\ref{fig:SpinEntg} for an illustration). Exactly how 
large the subsystems must be before they become entangled has been explored in the case of 
random states~\cite{Bhosale12}.

In the extreme case, the entanglement between two spins in a spin chain whose state obeys the volume law is 
typically zero. However if there is a transition to a localized phase where the ETH is violated, it can also 
be characterized by the appearance of entanglement among such small subsystems as described in 
Fig.~\ref{fig:SpinEntg}. Thus the purpose of this paper 
is to study the well-established measure of concurrence to measure the entanglement between two spin$-1/2$ 
particles when the whole system undergoes a transition to a MBL phase. Although several remarks about 
entanglement in the MBL phase were inferred from spin-correlations in the past, for 
example in Ref.~\onlinecite{pal10}, we are not aware of works that calculate the actual entanglement between 
two spins across the many-body localization transition. 

Moreover, concurrence is an entanglement monotone that has been measured experimentally in some contexts. 
In trapped atomic ion~\cite{Jurcevic2014} and cold atom~\cite{TakeshiBH2015} setup the concurrence dynamics 
after a global quench has been measured. Therefore understanding the behaviour of concurrence across the 
many-body localization transition will be useful for future experiments. 
 
\section{Model and method}
Consider a  model in which the transition to a MBL phase has been well-studied already~\cite{pal10},  the XXZ 
spin-chain:
\beq
H=\dfrac{1}{2}\left(\sum_{i=1}^{L-1} \sigma^x_i \sigma^x_{i+1}+\sigma^y_i \sigma^y_{i+1}+\Delta \sigma^z_{i} \sigma^z_{i+1}\right) +\sum_{i=1}^L h_i \sigma^z_i,
\label{XXZHam}
\eeq
where $h_i$ is a random magnetic field chosen from the uniform distribution on the interval $[-W,W]$ and 
$\Delta$ is an interaction strength. If $\Delta=0$ the model is equivalent to one of non-interacting 
fermions and even with the disordered magnetic field, it is integrable. The fluctuations of 
the energy levels is that of independent random variables and for example the unfolded nearest neighbor spacing 
distribution is Poisson ($e^{-s}$, where $s$ is the normalized nearest energy level spacing). When $\Delta$ 
is tuned away from zero, there is interaction in the 
underlying fermion model, and the nearest neighbor spacing for example could become that of the Gaussian 
Orthogonal Ensemble (GOE), that is of relevance to time-reversal symmetric complex quantum systems, and is characterized by level repulsion and spectral rigidity. The eigenstates in this phase have a volume law, and possess large multipartite entanglement.

However for a given interaction, if the disorder strength is sufficiently large, there is a transition from this ETH phase to the MBL phase and the Poisson level 
statistics reappears. This signals an interacting but localized phase, and is one of the first ways in which the MBL 
transition was studied. For the Hamiltonian defined in Eq.~\eqref{XXZHam} when $\Delta=1$, such a 
transition is believed 
to happen when $W$ exceeds $\approx 3.5$ in the large system size 
$L$ limit~\cite{pal10,Berkelbach:2010ib,DeLuca:2013ba,BarLev:2015co,luitz15,bera15}.

We will study the above mentioned model~(Eq.~\eqref{XXZHam}) using the exact diagonalization method at 
finite system sizes $L=10, 12, 14, 16$ and $18$ with open boundary condition. Typically for small system 
sizes upto $10^4$ disorder realizations are used to perform the disorder average and $500$ disorder 
realizations for $L=18$. Around 50 states are used from the middle of the spectrum to perform an additional 
average to decrease the statistical noise.

The MBL transition has been characterized in ways other than a transition in level statistics, mainly through various properties of the eigenstates. Entanglement has also been extensively studied in these transitions \cite{kjaell14,Grover:2014wm,luitz15} and fluctuation of the von Neumann entropy between two halves of the spin chain shows a maximum around the transition point. In all of these studies states are considered that are well away from either ends of the spectra: the ground state or the most excited 
state. It may be noted that a logarithmic growth of entanglement entropy in non-stationary states is also 
seen as a sign of interactions in such spin systems~\cite{Znidaric:2008cr, bardarson12,Serbyn:2013he}.

While such previous studies concentrate on the entanglement between two halves of the system, that may be 
characterized as macroscopic entanglement, this paper considers the complementary question of inter-spin 
entanglement. It is well-known that while the entanglement between any bipartition of a many-body system in a 
collective pure state can be measured as the von-Neumann entropy of any of the two subsystems, there is no 
such measure when the collective state is mixed. However the entanglement in the mixed state of  two spin-1/2 
particles is an exception, and concurrence~\cite{Wootters97,Wootters98} has long been used as such a 
measure. The entanglement of formation is a monotonic function of  concurrence, and it arises as a solution of 
a optimization problem. It has been used in numerous studies of spin chains and in particular derivatives of 
it can be an order parameter in low-temperature quantum phase transitions, such as in the XY and transverse 
Ising model \cite{Osterloh02}.

\begin{figure}[tbhp]
\centering
\includegraphics[width=0.999\columnwidth]{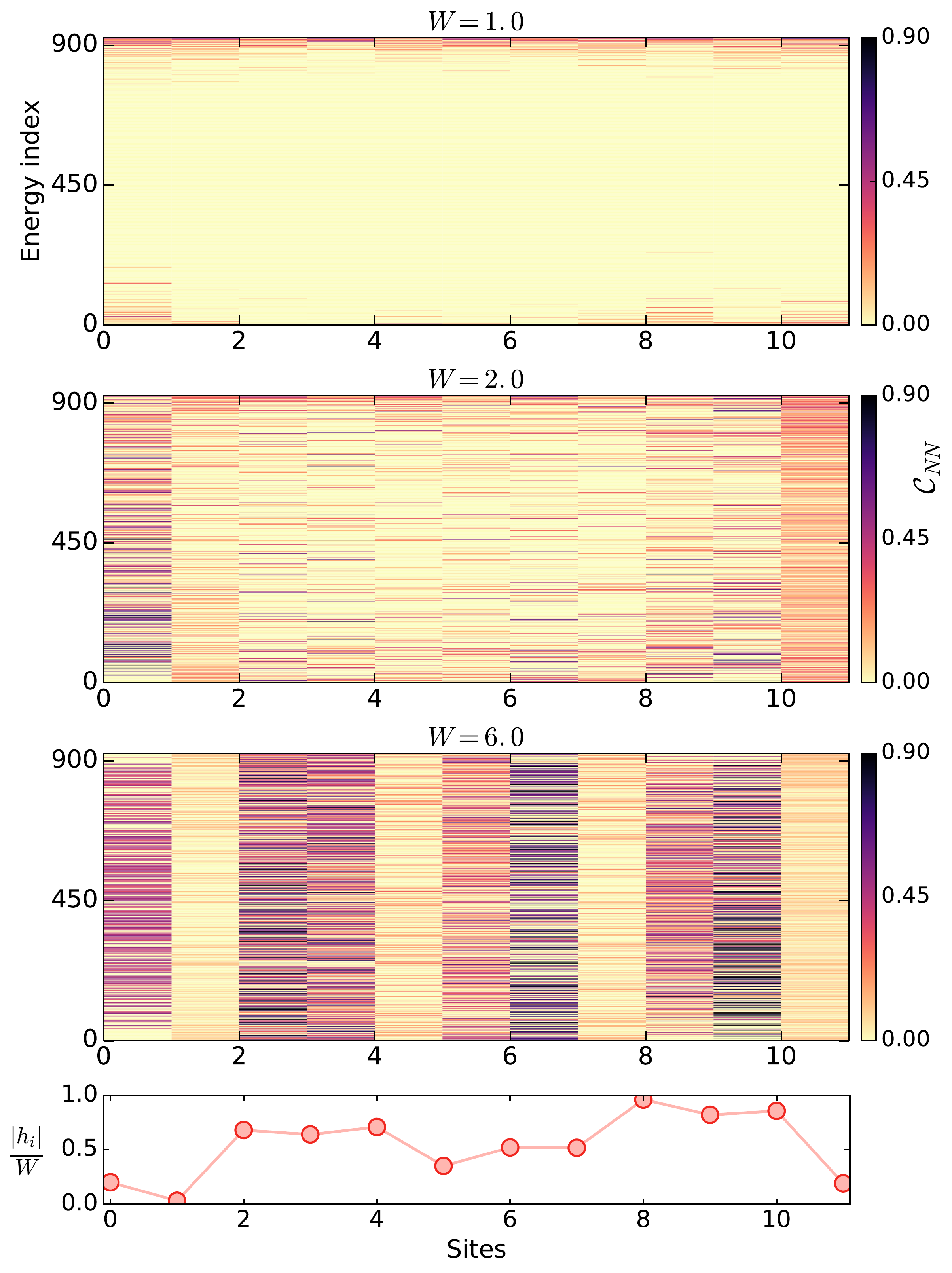}
\caption{The nearest-neighbor concurrence $\Cn_{NN}$ in eigenstates of the XXZ model is shown as the intensity for a 
given disorder configuration. The 
vertical axis is the state number, with the ground state being $0$. The horizontal axis has the $11$ nearest 
pairs with $L=12$ spins and  $\Delta=1$. The disorder strength $W=1.0$, $W=2.0$ and $W=6.0$ from top in the 
first three panels. The disorder configuration ($h_i$ when $W=1$) is identical in all cases and is shown in 
the last panel. This highlights that in the MBL phase the concurrence is correlated with disorder configuration and probe 
the resonances in the configuration. }
\label{fig:EnNNCW}
\end{figure}
\subsection{Concurrence as a measure of two body entanglement}
For completeness, concurrence is first defined in Ref.~\onlinecite{Wootters98}, for further details see 
also Ref.~\onlinecite{AmicoRMP08}. 
If $L$ spins are in a joint pure state $|\psi\kt$, 
let $\rho_{ij}$ be the state of two spins, one at site $i$ and the other at $j$.  It is in general a mixed state and the von 
Neumann entropy of the single spin reduced density matrices (RDM) does not have the interpretation of being 
the entanglement between the two spins. The state $\rho_{ij}$ is separable if there exists single spin density 
matrices $\rho_i^k$ and $\rho_j^k$ such that $\rho_{ij}=\sum_k p_k \, \rho_i^k \otimes \rho_j^k$, where $p_k$ 
are probabilities. If this is not possible, the mixed state is entangled. Thus for mixed states entanglement 
is a more subtle property than not being a product state. One of the central reasons for the difference is 
that mixed states can arise as mixtures of pure states in infinitely many physically different ways 
\cite{NielsenChuang}.

The set  of two-spin states $\{ |\phi_k \kt \}$ is a purification of $\rho_{ij}$ if 
$\rho_{ij}= \sum_k p_k |\phi_k \kt \br \phi_k|$, where $\{p_k\}$ are such that they are non-negative and add to $1$ 
(i.e. are probabilities). The entanglement of formation is defined as the minimum value of the average entanglement 
$\sum_k p_k {\cal E}(|\phi_k\kt)$, where ${\cal E}(|\phi_k\kt)$ is the entanglement of the pure 
bipartite state.
The minimization is over all the (infinite) possible purifications.  While this problem is hard in general, 
for 
two spin $1/2$ particles it was solved when it was shown that the concurrence as defined below is a monotonic function 
of the entanglement of formation. 

The concurrence in two qubits $i$ and $j$ that
are in the joint state $\rho_{ij}$ is given by the following procedure~\cite{Wootters98}:
Construct the matrix $\rho_{ij}\,\tilde{\rho}_{ij}$, where $\tilde{\rho}_{ij}
  =\, \sigma^y\otimes \sigma^y \rho_{ij}^*
\sigma^y\otimes \sigma^y$, and the complex conjugation is done in the standard computational
basis. Let the eigenvalues (guaranteed to be positive) be $\{\ld_1 \ge \ld_2 \ge \ld_3 \ge \ld_4\}$, then
  the concurrence is $\Cn_{i,j}=\mbox{max}\left (\sqrt{\ld_1}-
  \sqrt{\ld_2}-\sqrt{\ld_3}-\sqrt{\ld_4},\,0\right)$. 
  This is such that $0\le \Cn_{i,j}\le 1$, with the concurrence vanishing only for separable states and reaches unity for 
maximally
    entangled ones such as the singlet state. 
    
The Hamiltonian in Eq.~\eqref{XXZHam} commutes with the total spin $\sum_i \sigma_i^z$. Attention in this 
paper is restricted to the total spin $0$ sector. The 
conservation of total spin is a strong condition and in this case the non-zero elements of the  two-spin reduced density 
matrix $\rho_{ij}$ can be written in terms of spin-expectation values as
\beq
\begin{split}
& \br ab|\rho_{ij}|ab\kt = \df{1}{4}\br(1-(-1)^a\sigma_i^z) (1-(-1)^b\sigma_j^z)\kt, \;a,b \in \{0,1\},\\
&   \br 01|\rho_{ij}|10\kt =\br \sigma^{+}_i \sigma^{-}_j\kt,\,\br 10|\rho_{ij}|01\kt =\br \sigma^{-}_i \sigma^{+}_j\kt.
\end{split} 
\label{doublerdm}
\eeq
Here $\{|0\kt, \, |1\kt\}$ are eigenstates of $\sigma^z$ with 
eigenvalues $\pm1$ respectively and $\sigma^{\pm}= \frac{1}{2}(\sigma^x \pm i \sigma^y)$, or $\sigma^{+}=|0\kt 
\br 1|$, $\sigma^{-}=|1\kt \br 0|$. There are only $3$ real independent diagonal elements and one complex 
off-diagonal element in $\rho_{ij}$. The block-diagonal structure of such density matrices makes the 
concurrence in the state turn out to be of a simpler form~\cite{Connor01}:
\beq 
\Cn_{i,j}=2 \, \mbox{max}\{|z|-\sqrt{x y},\, 0\},
\eeq  
where $z=\br 01|\rho_{ij}|10 \kt$, $x=\br 00|\rho_{ij}|00 \kt$ and $y=\br 11|\rho_{ij}|11\kt$.

For definiteness we refer to states now as being composed of qubits or two-state particles, with the eigenstates 
labeled by $0$ and $1$ in the computational basis. The isomorphism with spin-1/2 states is evident. Monogamy 
of 
entanglement refers to the property that when two qubits are maximally entangled, then they cannot be entangled with any other 
qubits. This uniquely quantum correlation
also means that entanglement is shared in a many-body situation in many ways. 
It is known that in typical or random states of many qubits, multipartite entanglement is shared among many qubits 
rather than in a pairwise manner. For such states the probability that any two qubits are entangled is 
vanishingly small if the total number exceeds about $6$~\cite{Scott03}. In states that have further 
restrictions, for example 
restricted to a definite number of $0$ states, there could be enhanced concurrences. In \emph{one-particle 
states} that 
correspond to superpositions of one qubit in the $0$ state, with the rest in $1$ (or vice-versa), typical pairs of 
qubits can have fairly large entanglement with the concurrence decreasing as $1/L$~\cite{Lakshminarayan03}, 
where $L$ is the total number of 
qubits. For two-particle states it decreases as $1/L^2$, but for three and higher particle states it is exponentially 
small~\cite{Vijayaraghavan11} in $L$. Thus finding two qubit entanglement in a complex state, is rare.

\section{Results and Discussions}
In this section, we first describe the behavior of nearest neighbor concurrence in the many-body energy 
spectrum for a typical disorder realization. Next the scaling of average nearest neighbor concurrence is studied across 
the many-body localization transition. Finally we discuss the dependence of the concurrence on neighboring 
spin distances and system sizes in the localized phase. 
\subsection{Nearest neighbor concurrence across the many-body energy spectrum and resonances in disorder 
configuration}
\begin{figure}[tbp]
\centering
\includegraphics[width=0.986\columnwidth]{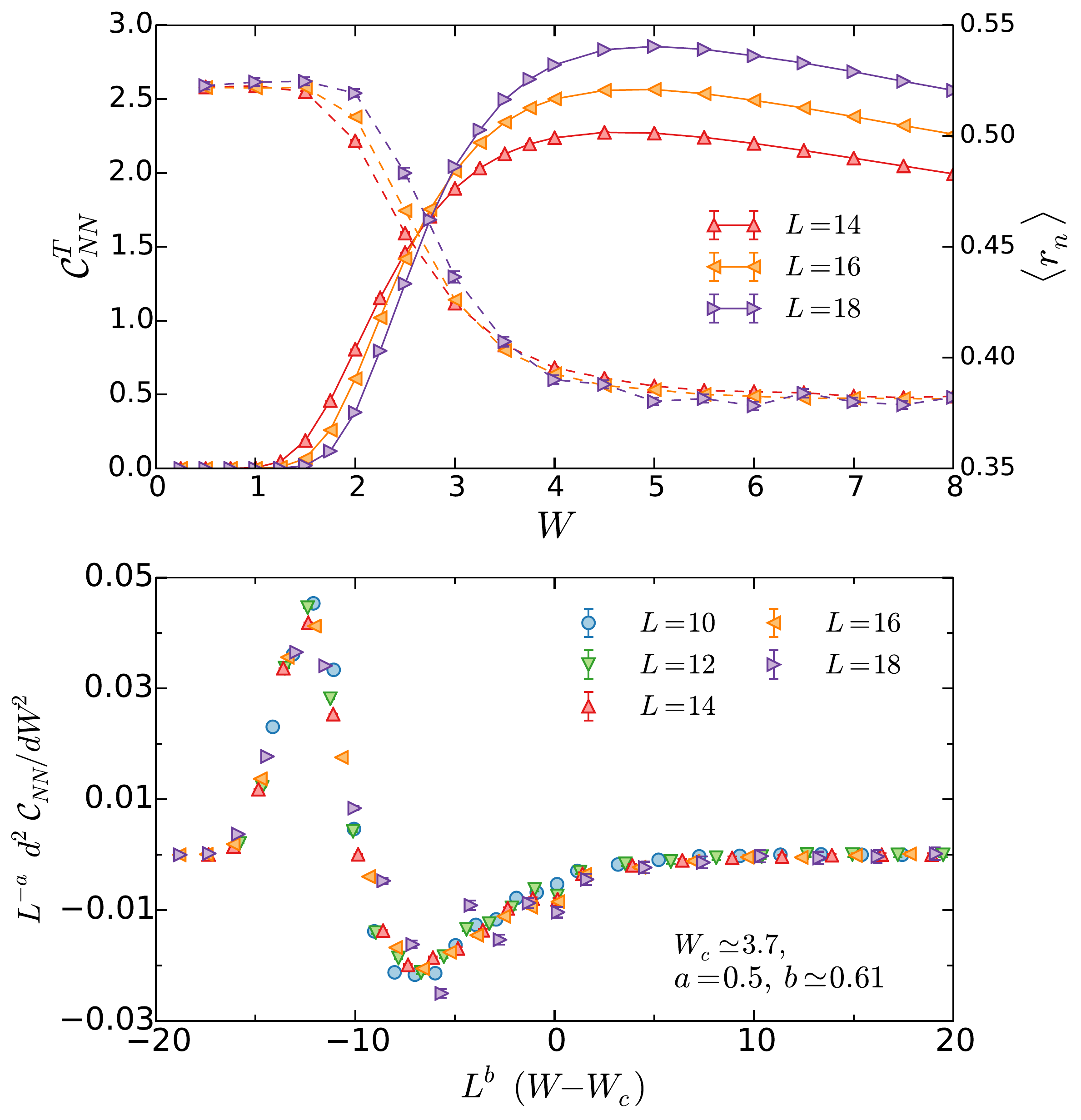}
%
\caption{Upper panel shows the the total nearest neighbor concurrence~(solid lines) and average of nearest energy level  
spacing ratios~(dotted lines) $\br r_n \kt$ for three different system sizes as a function of disorder strength for $\Delta=1.0$. 
It is clear that the many-body localization transition is captured similarly as in nearest energy level 
spacing and in concurrence. 
Lower panel shows a scaling collapse of the second derivative of the average nearest neighbor concurrence 
with respect to disorder.  From the scaling collapse critical disorder is estimated to be $W_c = 3.7 \pm 
0.2$.}
\label{fig:AvgTotvsW}
\end{figure}
\paragraph*{Typical concurrence} In Fig.~\ref{fig:EnNNCW} the nearest-neighbor concurrence is shown as a 
function of the energy across an entire spectrum for a specific disorder configuration and three different 
strengths. 
Top panel shows the concurrence for small disorder, while the middle panel for disorder values near the 
transition point $W=2$ and the lower panel for disorder value $W=6$, which is in the localized phase for 
system size $L=12$. The 
lowest panel shows the particular disorder configuration that is used in the calculation. For small disorder 
strength, but 
sufficiently far from the integrable value of $W=0$ the ETH or random phase dominates the spectra, except at 
the spectrum edges. It is clearly seen that in this case concurrence is present dominantly only in the lowest 
or highest 
excitations. The bulk of the states are free from nearest neighbor entanglement and in fact of any kind of two-body 
entanglement, suggesting that the other spins are acting as a thermal reservoir, and consistent with the fact that 
random matrix theory (RMT) fluctuations, particularly that of the Gaussian Orthogonal Ensemble (GOE) is obtained here. 
This also confirms that the ground state~(and the lowest excited states) is in the localized phase and the 
many-body localization transition is an excited state phase transition. 

Additionally it is also interesting to observe that the low lying excitations and the ground state have very 
large nearest neighbor entanglement. For translationally invariant states, under certain constraints, it was 
shown that the maximum possible nearest neighbor concurrence is about 0.434, and values close to this are 
indeed observed for the Heisenberg model~\cite{Connor01},
which corresponds here to $W=0$ and $\Delta=1$. It is therefore quite surprising to find such persistently 
large entanglement in the non-integrable case of $W=1$ when the bulk of the spectral fluctuations are that of 
RMT. The emphasis in this paper though will be on the excited states at the center of the spectrum or in the infinite temperature limit. 

\paragraph*{Resonances in disorder configuration} For higher values of disorder strength, it is clear from 
Fig.~\ref{fig:EnNNCW} that there is as strong variability in the concurrence across a spin chain for a given 
disorder configuration. In particular, it is interesting to note that in the localized phase across the energy 
spectrum large 
value of concurrence forms bands that is related to the disorder configuration.
The configuration is also shown at the bottom of the figure and it is quite clear that there is a correspondence: if $|h_{i+1}-h_i|$ is small ($\ll 1$) there is larger concurrence between spins $i$ and $i+1$. A heuristic understanding of 
this follows from the observation that for large $W$ the energy due to the external field may be considered 
the dominant one and the rest a perturbation. Concentrating on two neighboring spins $i$ and $i+1$, their 
unperturbed energies are $h_i+h_{i+1}$, $-h_i+h_{i+1}$, $h_i-h_{i+1}$ and $-h_i-h_{i+1}$, corresponding to the 
spin configurations  $00$, $10$, $01$ and $11$. When $h_i \approx h_{i+1}$ the perturbations can strongly  
couple  $01$ and $10$ to create entanglement (and concurrence) between the spins. In the opposite case when 
$h_i \approx -h_{i+1}$ it may seem that the configurations $00$ and $11$ may also couple to create  
entanglement, however as the perturbations preserve total spin, there is no matrix elements to create this, and hence 
the dominant concurrences occur essentially due to resonances in the disorder configuration.  

\paragraph*{Perturbative analysis} To be more concrete consider the extreme case of $L=2$. This is also 
relevant for larger $L$ when the configuration consists primarily  of the first $L/2$ spins in say $0$ and the 
other in $1$. The perturbation that mixes into this state when $W< \infty$ flips only the central two spins 
which then get entangled. For a state of the form $\alpha|01\kt +\beta |10\kt$ the concurrence is $2 |\alpha 
\beta|$. A simple calculation then gives that 
\beq
\Cn_{NN}= \frac{1}{\sqrt{1+(h_1-h_2)^2}},
\eeq
which shows that when $h_1=h_2$ the concurrence is the maximum possible and $NN$ stands for nearest-neighbor. 
Considering that $h_i$ are sampled 
uniformly in $[-W,W]$, the average concurrence is
\beq
\begin{split}
\br \Cn_{NN} \kt &= \frac{1}{2 W^2} \,\left(2 W \, \sinh^{-1}(2W)+1-\sqrt{1+4 W^2}\right) \\
&=\frac{1}{W}( \ln W+ \ln 4-1)+\mathcal{O}(1/W^2),
\end{split}
\eeq
showing how the concurrence vanishes when the disorder strength is increased to infinity. Numerical results are presented below for larger number of spins, but it is noteworthy that the $L=2$ averages reflect the rather large concurrence present in the MBL phase. For instance the average value of concurrence from the above formula is $0.32$ for $W=8$. As the number of spins increase this decreases as well, and the interplay of concurrence as a function of $L$ and $W$ is investigated as  a scaling relation.

\subsection{Average concurrence, finite size scaling and qualitative phase diagram}
\begin{figure}[tbp]
\centering
\includegraphics[width=0.95\columnwidth]{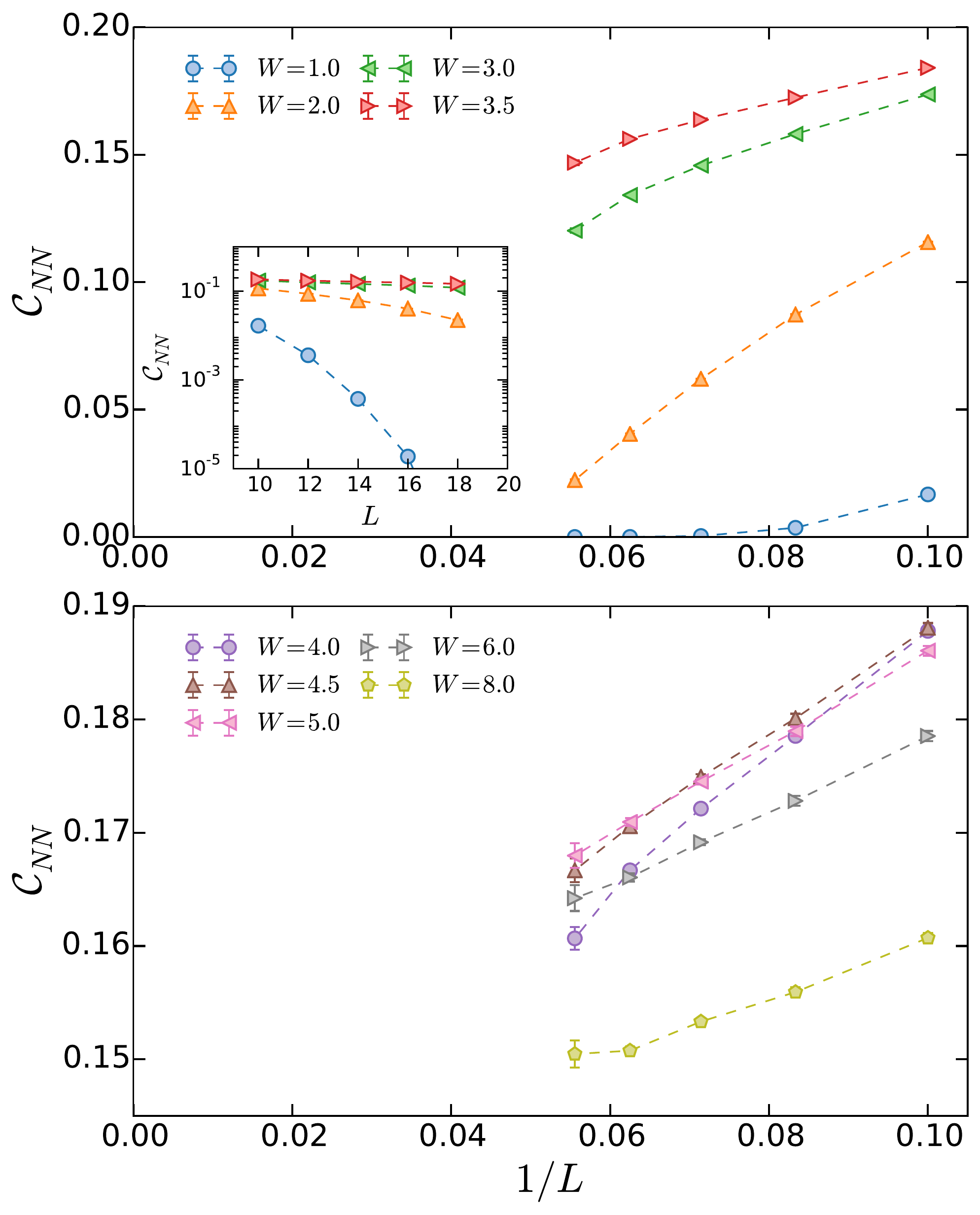}
    \caption{The nearest neighbor concurrence averaged over pairs and disorder configurations is shown as a 
function of (inverse) system sizes for various values of the disorder strengths, while 
the interaction $\Delta=1$. The concurrence for a fixed disorder strength seems to survive in the 
thermodynamic limit in the MBL phase~(lower panel), while it vanishes exponentially or faster in the ETH 
phase~(upper panel) as suggested by the inset of the top panel where the scale is linear-log.}
    \label{fig:ScalingwithL}
\end{figure}
\paragraph*{Average nearest-neighbor concurrence} In this section we will mainly concentrate on the 
average concurrence at finite energy densities. In particular, we concentrate at the middle of the energy 
spectrum, which corresponds to the infinite temperature limit in the thermodynamic system. Finally the energy 
dependence of 
the total concurrence will also be discussed.  Top panel of Fig.~\ref{fig:AvgTotvsW} shows the total nearest 
neighbor concurrence, 
\beq
C^T_{NN}=\sum_{i=1}^{L-1}  \Cn_{i,i+1},
\eeq 
averaged over disorder configurations for three  cases of $L=10,12$ and $14$. Shown in the same figure for convenience and comparison is the energy level average nearest 
neighbor {\it spacing} ratio, $\br r_n\kt$.  This spacing ratio is well-known to 
undergo a transition (after the initial integrable to GOE transition at $W=0$) from GOE to Poisson as the 
disorder strength $W$ increases. The Poisson value of $\br r_n \kt$ is $\approx 0.39$ while that of the GOE is $\approx 0.53$ \cite{Atas13}. 

\begin{figure}[tbp]
\centering
\includegraphics[width=0.95\columnwidth]{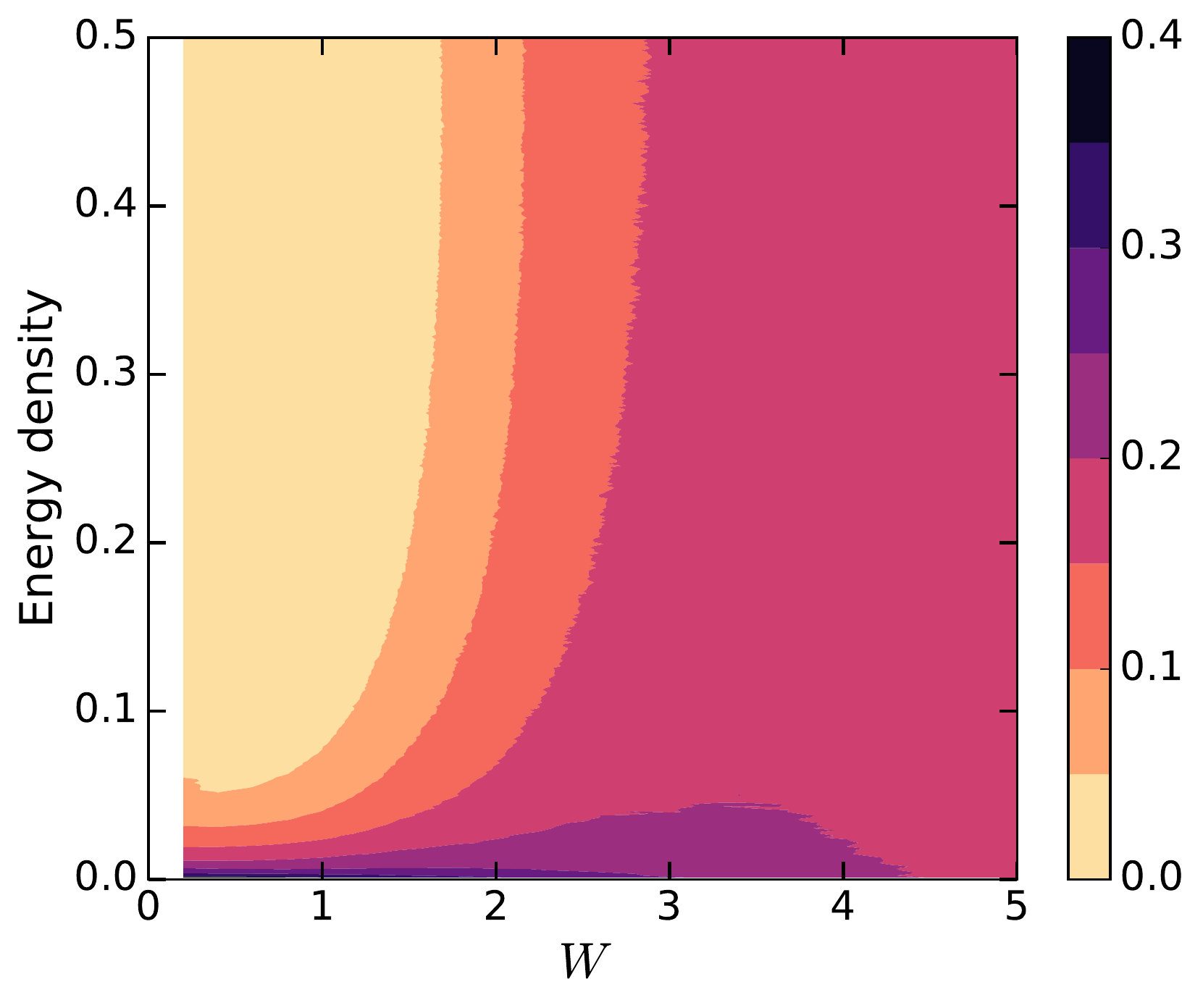}
    \caption{The nearest neighbor concurrence averaged over sites and disorder configurations is shown as a 
function of a normalized excitation level ($n/N$, where $n$ is the state's number and $N$ the number of 
states) and disorder strength. $L=12$ spins have been taken and averaged over $3 \times 10^3$ disorder 
realizations. Special nature of the low lying excitations are also visible as large concurrence.}
    \label{fig:EnrgWvsNNCL12}
\end{figure}
It is clear from Fig.~\ref{fig:AvgTotvsW} that when the nearest neighbor energy spacing distribution is that 
of the GOE concurrence is practically non-existent between any pair of neighboring spins. In fact no pair of 
spins are entangled. As soon as $\br r_n \kt$ starts to deviate from the GOE value, it also heralds the 
creation of two-spin entanglements as measured by the concurrence. Shown is simply the total of such 
entanglements between neighboring pairs of spins. It is seen that just after the RMT phase the total 
concurrence decreases with increasing system size, and we may expect that in the thermodynamic limit it even 
vanishes. However it is seen that the total concurrence curves cross each other at a value of $W$ that is 
roughly constant and occurs around the value where a transition to MBL is suggested to 
happen. Beyond this critical value of $W$, larger system sizes imply larger total concurrence. As will be 
presently shown most of the concurrence, when present, is in the nearest neighbors. It may be remarked that 
the average nearest neighbor concurrence per pair of spins $\Cn_{NN}^T/(L-1)$ decreases uniformly with $L$, 
across the entire range of $W$. At large disorder, though, the effect of finite size systems becomes more 
evident. In a sense that within these system sizes we observe that the concurrence decays as $1/W$, 
which is a result of a simple perturbation theory for large $W$. The very fact that the 
perturbation theory works suggest that the effect of interaction in this regime for this particular  
observable is negligible and the dominant physics is due to conventional Anderson localization rather the MBL, 
which is primarily due to interplay of interaction and disorder.

\paragraph*{Finite size scaling} In order to quantify the critical point of the phase transition, we look at 
the derivatives of the concurrence as a function of disorder. The lower panel of Fig.~\ref{fig:AvgTotvsW} 
shows a 
scaling collapse of second derivative of the concurrence for different values of systems sizes. We choose the 
following scaling form to perform the finite size scaling analysis
\eq{
  \frac{d^2 \Cn^T_{NN}}{d W^2} = L^a \Phi(L^b (W-W_c)), 
  }
where $\Phi$ is an unknown scaling function, which in practice is determined by the chi-square minimization 
procedure. In the lower panel of Fig.~\ref{fig:AvgTotvsW} the collapse is observed for the 
following parameters: $W_c = 3.7 \pm 0.2$, $a \simeq 0.5 $ and $b = 0.6 \pm 0.1$. The scatter in 
the $L=18$ data set is 
due to numerical noise and large statistical fluctuation. Such a uniform scaling collapse before and after the 
transition could be obtained for the second derivative, but not the concurrence or its first derivative. The 
critical value obtained is in agreement with those obtained by other measures~\cite{pal10, luitz15, bera15},  
although recently a larger critical disorder value in this model has also been predicted~\cite{Davakul15}.

\paragraph*{L-dependence} Fig.~\ref{fig:ScalingwithL} shows the average concurrence per pair of nearest 
neighbor spins as a function of system size $L$. As mentioned in the introduction random states are expected 
to have concurrence that is at least super-exponentially small in $L$. This is consistent with the data in the 
top panel of this figure for $W=1$, where except for small system sizes the probability of finding non-zero 
concurrence is negligible. For higher values of $W$, such as $3$, an exponential dependence seems plausible. 
In the transition region and in the deep MBL phase the behavior with systems size is very different showing a 
tendency for there to be entanglement in the large system size limit. This in turn reflects the 
quasi-integrable nature of the states in this phase.

\paragraph*{Qualitative phase diagram} Fig.~\ref{fig:EnrgWvsNNCL12} shows the average (over sites and 
disorder configurations) nearest neighbor 
concurrence,  as a function of energy density and disorder strength. A typical picture emerges 
showing that for a given disorder strength ETH behavior sets in at excited states. 
The qualitative pictures is consistent with earlier studies in a sense 
that at small energy density the states are localized thus the concurrence is finite. Whereas in the ETH 
phase the concurrence is strictly zero. In the middle of the spectrum the phase transition happens $W_c \ll 
3.7$ for system size $L=12$.
The special nature of the low-lying excitations are visible as those with large concurrence.

\subsection{Finite range of entanglement in the many-body localized phase}
\begin{figure}[tbp]
\centering
\includegraphics[width=0.975\columnwidth]{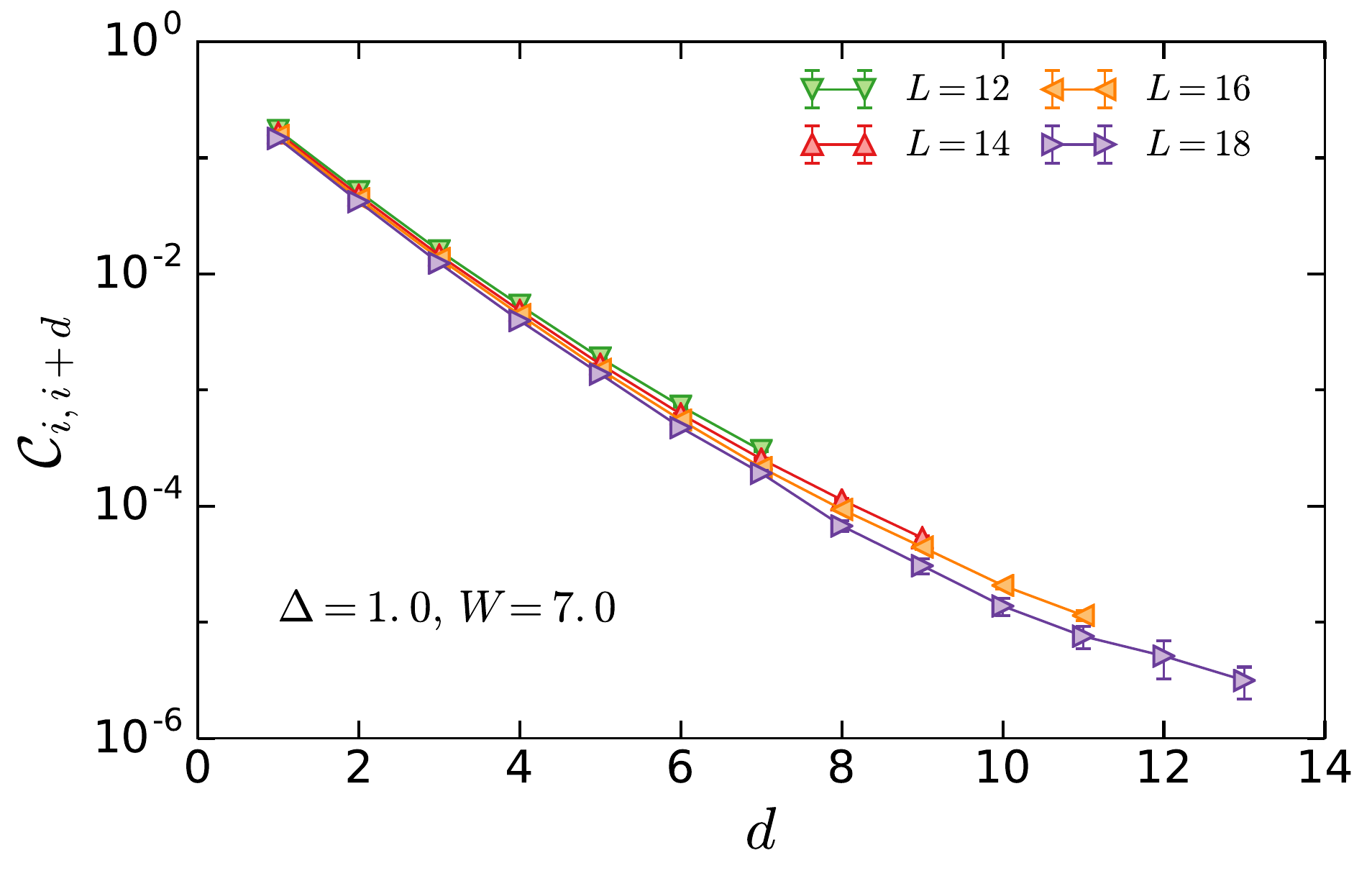}
    \caption{The concurrence averaged over sites and disorder configurations is shown as a 
function of distance for different systems sizes~($L=12, \ldots, 18$) in the MBL phase. 
The y-axis is in log scale to emphasize the exponential decay of average concurrence in the localized phase. 
The parameters are $W=7.0$ and $\Delta=1.0$. Note that in the ETH phase the concurrence is strictly zero 
across the chain length. }
    \label{fig:CvsddiffL}
\end{figure}
\paragraph*{Exponential decay of entanglement} To understand the range of entanglement in the localized phase 
we investigate the concurrence 
in a given system size $L$. Fig.~\ref{fig:CvsddiffL} shows the  entanglement between pairs of spins as a function 
of distance between them in a linear-log plot for disorder $W=7.0$ and interaction $\Delta=1.0$. The average 
of $\Cn_{i,i+d}$ is calculated for a given $d$ over different disorder realizations and sites $i$ and 
numerical data supports
\beq
\Cn_{i,i+d}\sim \Cn_{i,i+1}\exp(-d/\xi_E),
\label{eq:exp}
\eeq
that is at short distances $d \ll L$ the decay of concurrence is exponential. With increasing distance the 
concurrence as a function of distance has a slight curvature, which is related to finite size effects.  
It is interesting to note that the exponential decay of concurrence in the localized phase is reminiscent of 
exponential decay of correlation. However, in the ergodic phase the behaviour is completely reversed in a 
sense that spins are long range correlated but any two body entanglement~(even nearest neighbor) is absent. 
As mentioned above inferences about entanglement from correlation could be misleading.  

\begin{figure}[tbh]
\centering
\includegraphics[width=0.975\columnwidth]{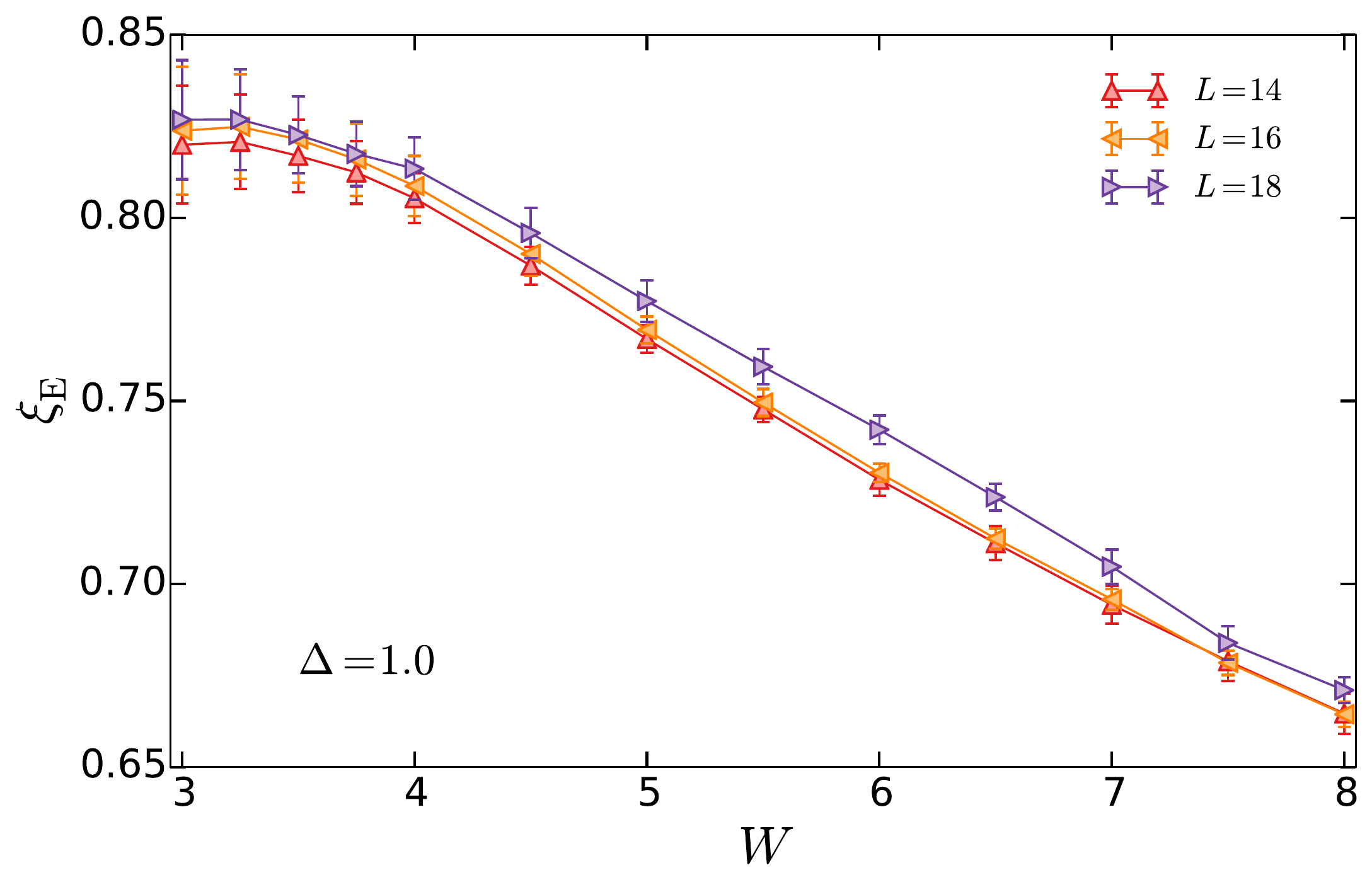}
    \caption{The entanglement length scale $\xi_E$ as function of the disorder for different systems 
sizes~($L=14, 16, 18$). The entanglement length is extracted by fitting the concurrence with 
Eq.~\eqref{eq:exp}. The saturation of the entanglement length close to the transition could be related to 
finite size effects. }
    \label{fig:LocLenvsW}
\end{figure}
\paragraph*{Entanglement length} Fig.~(\ref{fig:LocLenvsW}) shows the variation of the entanglement length 
$\xi_E$ with the disorder strength $W$ for various system sizes. It is interesting that in the range of 
parameters where the MBL transition happens (around $3-4$) there are substantial fluctuations in the 
entanglement length which is also large. As the MBL phase sets in the fluctuations are smaller and a more or 
less linear decrease in the localization length is observed.

\section{Conclusions}
In this work we proposed a way to investigate the local entanglement structure across the many-body 
localization transition by measuring the concurrence. First, we showed that the nearest neighbor concurrence, 
which is a measure of entanglement between two spins is intimately related to the disorder configuration. In 
particular, larger concurrence between two nearest neighbor spins corresponds to smaller difference between 
neighboring disorder configuration. This is clearly visible in Fig.~\ref{fig:EnNNCW} and consistent 
across the many body energy spectrum in the localized phase.

We further investigated the average concurrence at finite energy density and showed that in the RMT phase the 
concurrence vanishes exactly, implying all spins pair are disentangled. The entanglement is shared between 
all spins, which give rise to volume law bipartite entanglement entropy behaviour in an eigenstate.  In the 
MBL phase the concurrence is finite and the total nearest neighbor concurrence increases with increasing 
system sizes. At even stronger disorder~($W \gtrsim 6.5 $ for $\Delta=1$ and $L=18$) the concurrence decreases 
as $1/W$. The decrease can be understood from the perturbative analysis in the large disorder limit. The fact 
that in this limit the perturbative analysis is valid is an interesting observation. The validity of the 
perturbation theory strongly suggests that in this regime the effect of interaction is small and the 
physics is dominated by conventional Anderson localization rather than the many-body localization effect for 
these small system sizes for this observable. Eventually, in the thermodynamic limit, $L 
\rightarrow \infty$, one would expect that the decay of concurrence will be delayed and will be pushed to 
infinite disorder strength, such that the MBL phase will have a finite concurrence. 

The above fact is confirmed by analysing the system size dependence of the average nearest neighbor 
concurrence, $\Cn_{NN}$. It is found that in the RMT phase the concurrence decay super-exponentially with the 
system size, whereas in the MBL phase the $\Cn_{NN}$ is showing a tendency to survive in the thermodynamic 
limit. 

Next, to accurately calculate the phase transition we looked at the non-analyticity of the second derivative 
of the concurrence with respect to disorder, which is related to the variance. By performing a 
finite size scaling analysis we found the critical disorder strength to be around $W_c = 3.7 \pm 0.2 $, 
which is consistent with previously estimated value using other observables, including nearest energy level 
spacing.

Finally, we studied the distant spins concurrence in the MBL phase. We found that at small 
distance between spins the concurrence decay exponentially with the distance. The correlation length 
associated with the decay is also investigated and found to be saturating close to the transition point.

Several interesting directions of further research are opened due to the understanding of the concurrence in 
MBL phase. In particular, the dynamics of the concurrence after a quench is an immediate extension of this 
work. Furthermore, it would be fruitful to understand how much quantum entanglement is stored in local spins 
during unitary evolution in the localized phase, which could be used for storing quantum memory in future 
applications. Very recently the propagation of concurrence has been measured in a trapped atomic ion 
setup~\cite{Jurcevic2014} and in a Bose-Hubbard chain in an optical 
lattice~\cite{TakeshiBH2015}; thus studying the dynamics after a quench in the MBL phase will also 
be experimentally relevant.

\section{Acknowledgment}
We are grateful to Jens H Bardarson, Frank Pollmann  and Giuseppe De Tomasi for several 
insightful discussions and in particular, to Jens H Bardarson for collaboration on a related work. We would 
also like to thank MPIPKS visitor program for hospitality. 
\bibliography{references,refQIrmt}
\end{document}